\documentclass[12pt]{iopart}

\usepackage{iopams}
\expandafter\let\csname equation*\endcsname\relax
\expandafter\let\csname endequation*\endcsname\relax
\usepackage{amsmath}

\usepackage{graphicx}
\usepackage{color}
\usepackage{hyperref}
\usepackage{url}
\usepackage{tikz}
\usepackage{blkarray}
\usepackage{mathpazo}
\usepackage{cite}

\newcommand{\vect}[1]{ {\vec{ #1 }} }

\newcommand{\ket}[1]{\left| {#1} \right\rangle}

\newcommand{\braket}[3]{\left\langle {#1} \left| {#2} \right| {#3} \right\rangle}

\begin{document}

\title[Non-Lorentzian atomic natural line-shape of core level multiplets: Access ...]{Non-Lorentzian atomic natural line-shape of core level multiplets: Access high energy x-ray photons in electron capture nuclear decay.}

\author{Marc~Merstorf, Martin~Bra{\ss} and Maurits~W.~Haverkort}
 \address{Institute for Theoretical Physics, Heidelberg University, Philosophenweg 19, 69120 Heidelberg, Germany}

\date{\today}
\begin{abstract}

We present a method to calculate the natural line width and energy dependent line shape due to fluorescence decay of core excited atoms within a full relativistic multi-reference configuration interaction theory. The atomic absorption lines show a deviation from a Lorentzian line-shape due to energy dependent matrix elements of the localized electronic state coupling to the photon field. This gives rise to spectral lines with small but visible asymmetry. One generally finds an excess of spectral weight at the high energy shoulder of the atomic absorption line. We present the example of nuclear decay of $^{55}$Fe by electron capture of an inner-shell core electron. We show that the amount of ionizing radiation in the energy window between 50 and 200 keV is around one order of magnitude larger due to the energy dependent fluorescence yield lifetime compared to the value one would obtain if one assumes a constant fluorescence decay rate. This yields a total change of energy deposited into ionizing radiation of about 1\textperthousand. Our calculations are in good agreement with previous calculations and experimental observations where data is available. Our results can be further validated by high precision measurements of electron capture nuclear decay spectra using recently developed experimental methods.

\end{abstract}


\maketitle

Localized excited states in atoms, molecules and solids are no eigenstates of the Hamiltonian describing electrons, nuclei and photons. Due to the coupling of a localized state to a continuum of states the localized state will always have an energy spread. To a good approximation the energy dependent resonant line shape of atomic excitations is given by a Lorentzian function, related to a constant exponential decay of the localized excited state in time. This constant decay rate allows one to express the line-width or life-time of the excited state by a single number. 

Core excitations can, for example, be induced by the absorption or scattering of an x-ray photon. X-ray core level spectroscopy is a widely used technique to determine the physical and chemical properties of solids \cite{de2008core,glatzel2005high,yano2009x,DEGROOT2021147061}. Especially for resonant x-ray scattering experiments theoretical knowledge on the state dependent fluorescence lifetime is important to quantitatively determine the scattering cross section. Besides x-ray excitations one can also induce core excited states by nuclear decay via an electron capture event \cite{bergmann1999high,glatzel2001influence,Velte19,Ranitzsch20}. Independent of how the core excited state is created, one finds an energy dependent spectrum of excitation probabilities. 

We developed and implemented a method to calculate the fluorescence lifetime of core excited states and the resulting core excitation spectrum over a large energy window around the atomic resonance. The fluorescence line-broadening we find is not expressed by a single energy independent constant, but as an energy dependent function, i.e. the self-energy, $\Sigma(\omega)$. The real-part of the self-energy relates to an energy shift of the core excited state due to the coupling to the photon field, the imaginary part of the self-energy yields an energy dependent broadening of the excitation resonance. The $n$-electron plus nucleus many body problem is treated on a full relativistic configuration interaction level, starting from a basis of Kohn-Sham orbitals determined by density functional theory \cite{engel2011density,koepernik1999full,opahle1999full}, see \ref{sec:SolvingHAtom} for detail. Our method is not restricted to closed shell systems, but can treat any open shell system including the many-body multiplet splittings. The calculations are done and implemented in the many-body script language \textsc{Quanty} \cite{haverkort2014bands,haverkort2016quanty,brass2020ab}. 

We tested our method by calculating the electron capture spectrum of $^{55}$Fe. For the $K$-edge, a state with one hole in the $1 s$ orbital, we find a very strong energy dependence of the fluorescence induced self-energy, related to the energy dependence of the light matter interaction matrix elements. As a result we find about an order of magnitude larger number of decay events producing high energy photons. This excess of high energy photons is often referred to in the literature as internal Bremsstrahlung, related to a semi classical Bohr model like interpretation of atomic excitations.  The integrated amount of high energy excessive radiation observed experimentally \cite{isaac1990internal} and previously predicted theoretically \cite{glauber1956radiative,martin1958relativistic} is in good agreement with the results we find.

Several unstable nuclei can decay by capturing one of the inner shell electrons into the nucleus. In this process one of the atomic core protons is transformed to a neutron, thereby releasing a large amount of energy, $Q$. Within the electron capture process the atomic number is reduced by one, i.e. the element changes. The produced daughter atom is in an electronically excited state, whilst it has a hole in one of its inner shells and an additional valence electron compared to the ground-state of the daughter atom. Specific to electron capture and its inverse process of beta nuclear decay is that the released nuclear energy $Q$ is shared between a neutrino and the electronic excitations. Where the neutrino can move freely without interacting or damaging its surrounding, the energy deposited in the electronic sector is released largely via ionizing radiation in the form of electrons, and photons. The fraction of decay processes that happen with a specific ratio between energy deposited into the neutrino and the other degrees of freedom is known as the decay spectrum or the differential decay rate $\frac{\mathrm{d}\Gamma}{\mathrm{d}\omega}$. For an electron capture decay the differential decay rate depends on the matrix elements to capture an electron from one of the atomic basis orbitals as well as the number of many body final states available for the decay. The matrix elements are given by the weak interaction and the overlap between the electronic states and the nuclear wave function. The many body density of states depends on the neutrino density of states on one hand and the many electron density of states including a coupling to free electrons and photons on the other hand. 

There is an ongoing endeavour to measure and calculate the beta emission \cite{loidl2018metrobeta,loidl2019beta,kossert2022high,mougeot2014consistent,mougeot2015reliability,mougeot2015erratum} or the electron capture spectra \cite{faessler2014electron,faessler2015improved,faessler2015determination,faessler2017neutrino,croce2016development,de2016calorimetric,Velte19,Ranitzsch20,ge2021dy,friedrich2021limits} of several nuclear isotopes with high precision. These determinations are not only important from a metrology point of view \cite{pomme2016model}, but find applications in a variety of fields.  Using radioactive isotopes to determine the age of our solar system \cite{huss2009stellar,jorg2012precise} or for cancer treatment \cite{lee2012atomic,bavelaar2018subcellular,ku2019auger} require the detailed knowledge on the amount and energy of ionizing radiation produced during a nuclear decay. A precise determination of the nuclear decay spectra can be used to derive the neutrino mass \cite{kraus2005final,aseev2011upper,katrin2022direct,esfahani2017determining,alpert2015holmes,gastaldo2017electron,Velte19} or to rule out physics proposals beyond the standard model \cite{bezrukov2007searching,boyarsky2019sterile,smith2019proposed}. Furthermore a detailed knowledge on the energy, characteristics and amount of ionizing radiation produced by a nuclear decay is needed to calibrate liquid scintillation counting \cite{broda2007radionuclide,kossert2015importance}, one of the primary techniques to determine the activity of a radioactive sample. In order to understand the nuclear decay spectra in terms of atomic transitions detailed calculations are needed that include the full atomic detail of the decaying atom \cite{MOUGEOT2019108884, brass2018ab, Martins20, Martins20a, brass2020ab}. 

Due to the weak interaction a proton from within the nucleus of $^{55}\mathrm{Fe}$ can capture one of the inner shell electrons of the atom to form the nucleus of an  $^{55}\mathrm{Mn}^*$ atom and an electron neutrino $\nu_e$. The total energy released in the nuclear decay $Q=231$ keV \cite{chu1999lund} is shared between the electronic excitations of the  $^{55}\mathrm{Mn}$ atom and the energy of the electron neutrino. The differential decay rate $\frac{\mathrm{d}\Gamma}{\mathrm{d}\omega}$ presents the rate of decay as a function of the energy of the electronic excitations of the $^{55}\mathrm{Mn}$ atom. Using time dependent perturbation theory we find to first order in the weak interaction \cite{brass2018ab, brass2020ab}
\begin{eqnarray}
\label{eq:fracDecRateGeneral} 
\nonumber \frac{\mathrm{d}\Gamma}{\mathrm{d}\omega} \propto   -\mathrm{Im} & \left( D_{\nu}    \braket{\psi_0^{^{55}\mathrm{Fe}}}{\mathbf{T}^{\dag} \frac{1}{\omega- \mathbf{H} +E_0 +\mathrm{i}0^+}\mathbf{T}}{\psi_0^{^{55}\mathrm{Fe}}}  \right. \\ 
&  \quad \, \left.  -  \braket{\psi_0^{^{55}\mathrm{Fe}}}{\mathbf{T}^{\dag} \frac{1}{\omega + \mathbf{H} -E_0 +\mathrm{i}0^+}\mathbf{T} }{\psi_0^{^{55}\mathrm{Fe}}} \right),
\end{eqnarray}
with $\ket{\psi_0^{^{55}\mathrm{Fe}}}$ the ground-state of the $^{55}\mathrm{Fe}$ atom, $\mathbf{T}$ the operator representing the weak interaction describing the process of annihilating an electron, creating an electron neutrino and changing the $^{55}\mathrm{Fe}$ nucleus into a $^{55}\mathrm{Mn}$ nucleus \cite{brass2018}, $E_0$ the total ground-state energy of the $^{55}\mathrm{Fe}$ atom and $D_{\nu} =  \sum_{i=1}^3 |U_{ei}|^2 (Q-\omega)\sqrt{(Q-\omega)^2-m^2_{\nu_i}}$ the neutrino phase space factor with $m_{\nu_i}$ the neutrino mass of the $i$th neutrino mass eigenstate and $U_{ei}$ the Pontecorvo-Maki-Nakagawa-Sakata matrix \cite{Maki1962} describing the relation between the electron neutrino and the mass eigenstates of the neutrino. The Hamiltonian $\mathbf{H}$ includes all interactions, except the weak interaction, which is treated perturbatively, nor the kinetic and mass energy of the neutrino. The Hamiltonian $\mathbf{H}$ contains the full relativistic rest mass and kinetic energy of the nucleus and the 25 electrons. It furthermore includes the Coulomb interactions between the electrons and the interaction between the Mn nucleus and the electrons. On top of this, the Hamiltonian $\mathbf{H}$ includes the interactions between the electrons and the electromagnetic field, $\mathbf{H}_{e\gamma}$. The latter interaction allows the system to spontaneously emit a photon. Within the calculations the ground-state energy difference of $^{55}\mathrm{Fe}$ and $^{55}\mathrm{Mn}$ is set to the experimentally obtained parameter $Q$ such that no nuclear structure calculations are needed. The Hilbert space $\mathcal{H}$ for the Hamiltonian $\mathbf{H}$ is given by a sum over several subspaces $\mathcal{H}^{(n)}$ with constant number of $n$ photons, i.e. as a sum over states with zero photons $\mathcal{H}^{(0)}$ or one photon $\mathcal{H}^{(1)}$ or two photons $\mathcal{H}^{(2)}$, etc. 
\begin{eqnarray}
\mathcal{H} = \mathcal{H}^{(0)} \oplus \mathcal{H}^{(1)} \oplus \mathcal{H}^{(2)} \oplus \hdots.
\end{eqnarray}
The Hilbert space for states with $n$ photons is given by the tensor product of the Hilbert space of all electronic states of  $^{55}\mathrm{Mn}$ given as $\mathcal{H}_{^{55}\mathrm{Mn}}$ times the Hilbert space for a photon $\mathcal{H}_{\gamma}$ to the $n$-th power
\begin{eqnarray}
\mathcal{H}^{(n)} & = \mathcal{H}_{^{55}\mathrm{Mn}} \left( \otimes \mathcal{H}_{\gamma} \right)^n .
\end{eqnarray}
The Hamiltonian coupling states in sub-Hilbert spaces with one photon number different, i.e. $\mathcal{H}^{(n)}$ and $\mathcal{H}^{(n+1)}$, is given by the electromagnetic interaction operator $\mathbf{H}_{e\gamma} \approx \vect{\mathbf{p}} \cdot \vect{\mathbf{A}}$, with $\vect{\mathbf{p}}$ the electron momentum operator and operator $\vec{\mathbf{A}}$ creating or annihilating a photon. The interaction within a Hilbert subspace with constant photon number is given by the electronic Hamiltonian plus the energy of the photons. 

Diagonalizing the Hamiltonian $\mathbf{H}$ would lead to a continuous spectrum of eigenenergies $\omega_f$ and eigenstates $\ket{\psi_{\omega_f}}$. The operator $\mathbf{H}$ includes the interaction between the electrons and photons. As such the eigenstates are mixed states with different photon occupation numbers and photon energies. On a basis of the eigenstates of $\mathbf{H}$, with $\mathbf{H} \ket{\psi_{\omega_f}} = \omega_f \ket{\psi_{\omega_f}}$, equation \ref{eq:fracDecRateGeneral} for the differential decay rate becomes
\begin{eqnarray}
\frac{\mathrm{d}\Gamma}{\mathrm{d}\omega} \propto D_{\nu} \int_{0}^{\infty} \left|  \braket{\psi_{\omega_f}}{\mathbf{T}^{\phantom{\dag}}}{\psi_0^{^{55}\mathrm{Fe}}} \right|^2 \delta(\omega - \omega_f +E_0) \mathrm{d} \omega_f.
\end{eqnarray}
Calculating this integral would require to find all eigenstates of an infinite dimensional Hamiltonian that contains mixed states in terms of photon occupation number. 

\begin{figure}[t]
\includegraphics[width=\textwidth]{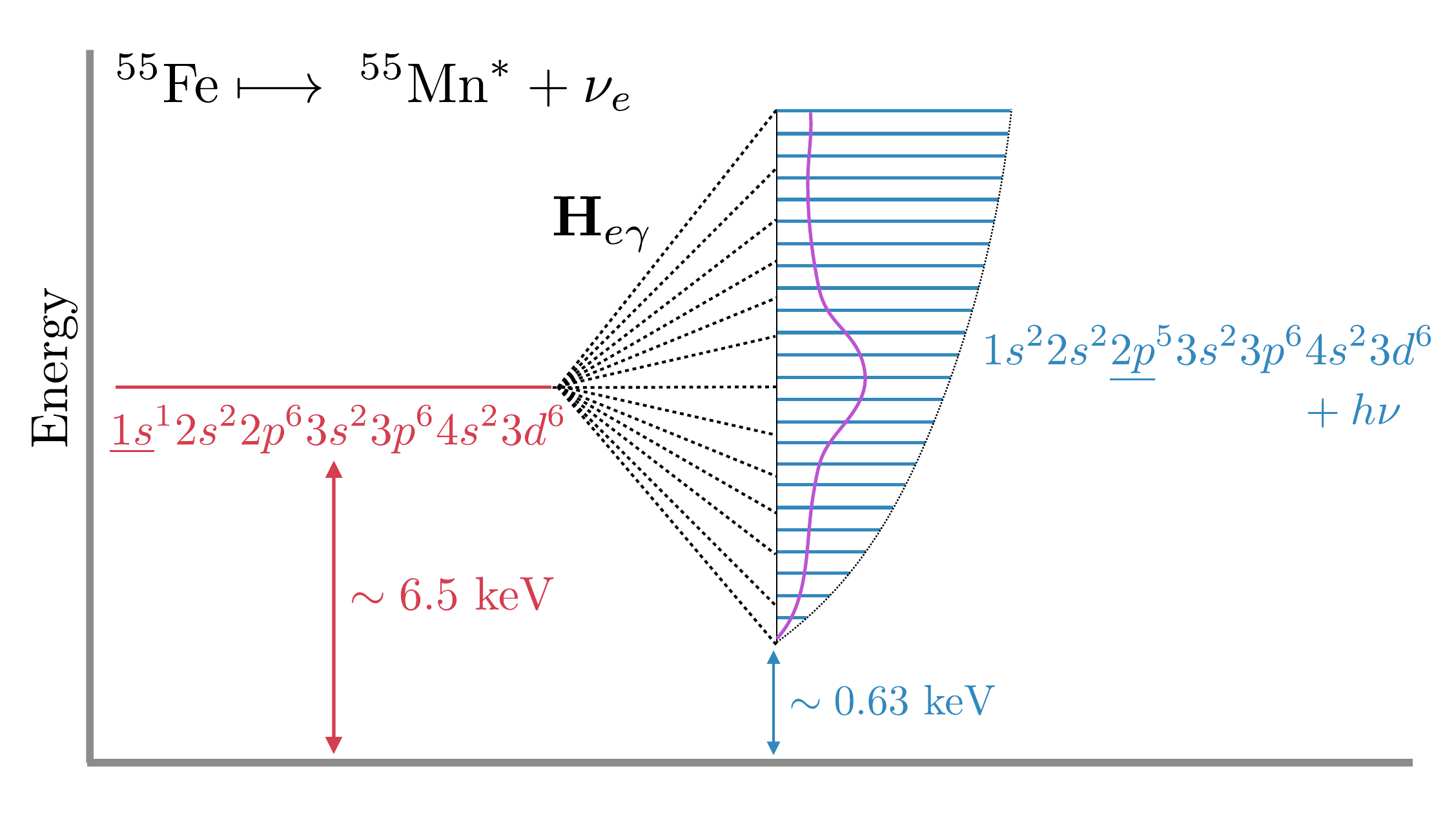}
\caption{Sketch following a specific decay channel for the nuclear decay of an $^{55}\mathrm{Fe}$ atom by electron capture. The energy is labeled with respect to the ground-state energy of the $^{55}\mathrm{Mn}$ atom. At $\approx$ 6.5 keV excitation energy (red line) one finds a state with one $1s$ electron hole and one valence $3d$ electron extra. This state is not an eigenstate of the Hamiltonian. For example, the state with a $1s$ core hole (red) can interact with states containing a $2p$ core hole and one additional photon (blue) via the electromagnetic interaction ($H_{e\gamma}$ black). The states with a $2p$ core hole and one addition photon start at $\approx 0.63$ keV, the excitation energy of a $2p$ core hole and form a continuum due to all possible photon energies. 
\label{fig:sketch}}
\end{figure}

In order to simplify the problem we can use that the state $\mathbf{T} \ket{\psi_0^{^{55}\mathrm{Fe}}} $ contains no free photons. The state $\mathbf{T} \ket{\psi_0^{^{55}\mathrm{Fe}}} $ thus is an element of the subspace $\mathcal{H}^{(0)}$. In matrix form on a basis of the subspaces $\mathcal{H}^{(0)}$, $\mathcal{H}^{(1)}$ etc. we can write for the Hamiltonian
\begin{eqnarray}
H =    \begin{pmatrix} 
      H^{(00)} & H^{(01)}_{e \gamma} & 0 & 0  \\
      H^{(10)}_{e \gamma} & H^{(11)} & H^{(12)}_{e \gamma} & 0  \\
      0 & H^{(21)}_{e \gamma} & H^{(22)} & \ddots   \\
      0 & 0 & \ddots & \ddots 
   \end{pmatrix}.
\end{eqnarray}
For the differential decay rate we find, with the help of Schur's formula,
\begin{eqnarray}
\label{eq:fracDecRate}
\nonumber \frac{\mathrm{d}\Gamma}{\mathrm{d}\omega}  \propto \, D_{\nu} \, \mathrm{Im} & \left(  \,\, \braket{\psi_0^{^{55}\mathrm{Fe}}}{\mathbf{T}^{\dag} \frac{1}{\omega- \mathbf{H}^{(00)} +E_0  - \mathbf{\Sigma}^+(\omega) }\mathbf{T}}{\psi_0^{^{55}\mathrm{Fe}}}  \right.  \\ &  \left. -  \braket{\psi_0^{^{55}\mathrm{Fe}}}{\mathbf{T}^{\dag} \frac{1}{\omega + \mathbf{H}^{(00)} -E_0 - \mathbf{\Sigma}^-(\omega)}\mathbf{T} }{\psi_0^{^{55}\mathrm{Fe}}} \right),
\end{eqnarray}
with the self-energy operator
\begin{eqnarray}
\label{eq:SelfEnergy}
 \mathbf{\Sigma}^{\pm}(\omega) = \pm\mathbf{H}^{(01)}_{e\gamma} \frac{1}{\omega \mp \mathbf{H}^{(11)} \pm E_0 \mp  \mathbf{H}^{(12)}_{e\gamma} \frac{1}{\omega \pm \mathbf{H}^{(22)} \pm \hdots} \mathbf{H}^{(21)}_{e\gamma} } \mathbf{H}^{(10)}_{e\gamma}.
\end{eqnarray}
The operator $\mathbf{\Sigma}(\omega) $ describes the process where a core excited state of  $^{55}\mathrm{Mn}$ decays by emitting a photon. The decay is not necessarily to the ground state of the Mn atom. The operator $\mathbf{H}^{(12)}_{e\gamma} \frac{1}{\omega - \mathbf{H}^{(22)} - \hdots} \mathbf{H}^{(21)}_{e\gamma}$ describes the process where a second photon is emitted. 

In figure \ref{fig:sketch} we illustrate a specific path for the decay. Within sub-Hilbert space $\mathcal{H}^{0}$ the operator $\mathbf{T}$ in equation \ref{eq:fracDecRate} can, for example, annihilate a $1s$ core electron, create a neutrino and change the Fe nucleus into a Mn nucleus. Within sub-Hilbert space $\mathcal{H}^{0}$ the resulting state has an electronic excitation energy of about 6.5 keV above the electronic ground state energy of a Mn atom. The neutrino energy must not be $Q-6.5$ keV, but can be any value, since the state $\mathbf{T} \ket{\psi_0^{^{55}\mathrm{Fe}}} $ is not an eigenstate of the Hamiltonian. The state can, for example, interact via the Hamiltonian $\mathbf{H}_{e\gamma}$ with an electronic state with a core hole in the $2p$ orbital and an additional photon. The photon can have any positive energy, such that there is a continuum of states with a $2p$ core hole and a photon, as indicated by the blue density of states in figure \ref{eq:fracDecRate}. The electromagnetic interaction results in a set of mixed states in terms of photon occupation number. Each state with a $2p$ core hole and photon gets an infinitesimal small admixture of the state with a $1s$ core hole without photon, as indicated in purple in figure \ref{eq:fracDecRate}. These mixed states form a continuum.

For a specific decay channel we can express the self-energy operator as a function. For example, the process depicted in figure \ref{fig:sketch} yields the self-energy for the state with a $1s$ core hole due to the fluorescence decay of a $2p$ electron into the $1s$ subshell,
\begin{eqnarray}
\label{eq:SelfEnergy2pto1s}
\Sigma_{2p,1s}(\omega) = \int_0^{\infty} \braket{\psi_{\underline{1s}}} { \mathbf{H}^{\omega_{\gamma}}_{e\gamma} \frac{1}{\omega - \omega_{\underline{2p}} - \omega_{\gamma}+ \mathrm{i}0^+ } \mathbf{H}^{\omega_{\gamma}}_{e\gamma}} {\psi_{\underline{1s}}} \mathrm{d} \omega_{\gamma}.
\end{eqnarray}
The operator $\mathbf{H}^{\omega_{\gamma}}_{e\gamma}$ depends on the electromagnetic field via the vector potential $\vect{\mathbf{A}} \sim e^{\mathrm{i} \vect{k} \cdot \vect{r}}$, with $\vect{k}$ the wave vector of the photon and $\hbar \omega_{\gamma} = \hbar c |\vect{k}|$ the photon energy. The matrix elements for operator $\mathbf{H}^{\omega_{\gamma}}_{e\gamma}$ between two bound electron orbitals scale as the Fourier transform of the product of these orbitals. These matrix elements are strongly wave vector and thus energy dependent.  \ref{sec:MultipoleExpansionLightMatter} contains  the details on how to calculate these matrix elements. 

\begin{figure}[htb]
\includegraphics[width=\textwidth]{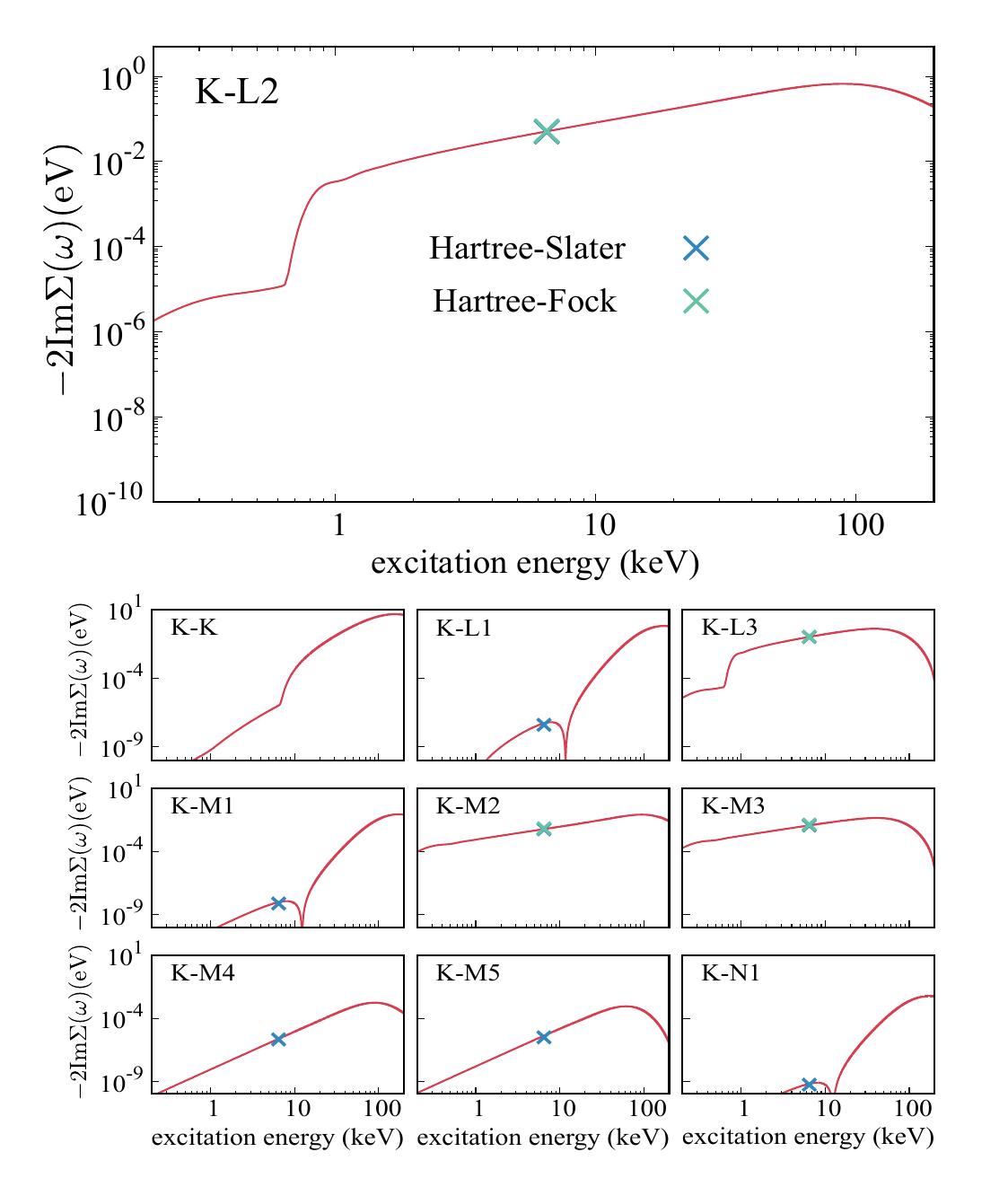}
\vspace{-0.75cm}
\caption{Energy dependence of the imaginary part of the self-energy for different fluorescence decay channels after the electron capture of a $1s$ electron in $^{55}\mathrm{Fe}$. Blue crosses Hartree-Slater results from reference \cite{scofield1974relativistic}. Green crosses Hartree-Fock results from reference \cite{scofield1974exchange}. 
\label{fig:selfenergyselectedchannels}
}
\end{figure}

In figure \ref{fig:selfenergyselectedchannels} we show twice the self-energy for several channels starting with the capture of a $1s$ core electron in $^{55}\mathrm{Fe}$. We show the imaginary part of the self-energy, which is the inverse of the life-time of the state, or the full width half maximum broadening. There is a strong energy dependence of the self-energy. The self-energy changes by several orders of magnitude. Note the logarithmic vertical scale. On resonance we can compare our calculations with calculations one can find in the literature. For these values, indicated by blue \cite{scofield1974relativistic} and green \cite{scofield1974exchange} crosses in figure \ref{fig:selfenergyselectedchannels} we find good agreement. We particularly find a strong increase of the self-energy above the resonant energy related to the binding energies of the core electrons. This is related to a strong increase in the transition matrix element due to magnetic dipole transitions.

\begin{figure}[htb]
\begin{center}
\includegraphics[width=\textwidth]{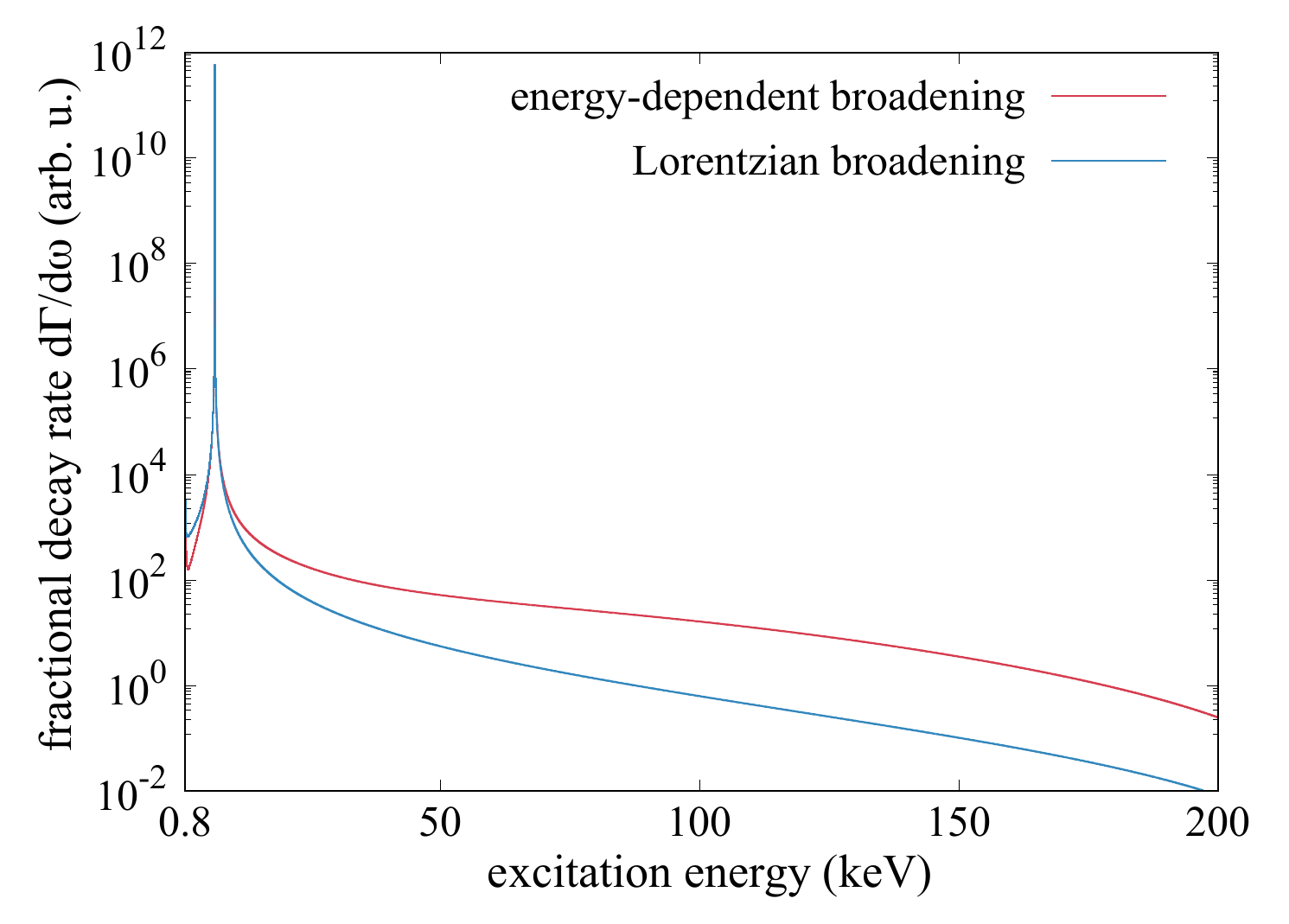}
\vspace{-1cm}
\caption{Electron capture spectrum of $^{55}\mathrm{Fe}$. Broadening due to fluorescence decay is represented by a constant self-energy (blue)  or an energy dependent one (red). The energy dependent matrix elements of Fluorescence decay lead to an increase of more than one order of magnitude in the ionizing radiation with energies in the range from 50 to 200 keV.  \label{fig:FeECSpectrum}}
\end{center}
\end{figure}

The self-energy of a core excited atomic state due to fluorescence decay is strongly energy dependent, over an energy window of several keV. This does not imply that the differential electron capture decay rate as a function of electronic excitation energy $\omega$ completely deviates from a Lorentzian line shape. Form experiments involving core level excitations it is well known that the line shape is within some approximation Lorentzian-like. The imaginary part of the self-energy at the resonance is of the order of several eV. The slope of the self-energy on the eV scale is small enough that one, at the resonance can approximate the line shape by a Lorentzian. 

In figure \ref{fig:FeECSpectrum} we show the differential electron capture rate as a function of $\omega$. With $Q-\omega$ the energy of the neutrino and $\omega$ the energy deposited into electronic excitations, approximately equal to the energy of the ionizing radiation released in an electron capture event. In blue we show the spectrum assuming a constant self-energy. In red we show the spectrum resulting from the energy dependent self-energy. For a window of several 100 eV around the resonance the difference between the two theories is modest. However, in the range of 50 to 200 keV where high energy ionizing radiation is emitted, we find an increase of events by more than one order of magnitude when we consider the energy dependent self-energy. Having 10 times more events whereby there is a large amount of energy released into ionizing radiation also changes the total amount of energy released into the ionizing radiation. The total energy released into ionizing radiation is roughly given by the first moment of the differential decay rate, i.e. the integral $\int_{0}^{\infty} \omega \frac{\mathrm{d} \Gamma}{\mathrm{d}\omega} \mathrm{d}\omega$. As most events happen close to the resonance, we find a change of about one per mille for the total energy released into the ionizing radiation. 

Our results are in excellent agreement with experimental results \cite{isaac1990internal} and previous mean-field calculations \cite{glauber1956radiative,martin1958relativistic} of the high energy excess of gamma rays produced after an electron capture event in $^{55}$Fe.  We find a fraction of $3.06 \times 10^{-5}$ events of the total to decay in the energy window from 35 to 231 keV. Which compares well to the experimental result of  $3.24 (6) \times 10^{-5}$ events \cite{isaac1990internal} and predictions by Glauber and Martin of $3.20 \times 10^{-5}$ events \cite{glauber1956radiative,martin1958relativistic}. 

In conclusion we present a method to include the energy dependent fluorescence lifetime of core excited states into many-body configuration interaction calculations of atomic excitations. The fluorescence decay generates a state and energy dependent self-energy matrix that can be used to calculate the broadening of core level excitations as one observes them in x-ray absorption, resonant inelastic x-ray scattering or nuclear decay by electron capture spectroscopy. We present calculations of the electron capture spectra of $^{55}$Fe. We find an excess of high energy decay events due to the energy dependence of the fluorescence decay matrix elements, in good agreement with experimental and previous mean-field theoretical predictions. Our predictions can be further validated by a comparison of the theoretical line-shape to high precision measurements of the electron capture spectra using modern metallic magnetic calorimeters \cite{Ranitzsch20}.

This work has been funded by the DFG (German Research Foundation) – Project-ID 273811115 – SFB 1225 ISOQUANT and Project-ID 449872909 – FOR 5249 QUAST. Part of this work was performed within the European project MetroMMC, “Measurement of fundamental nuclear decay data using metallic magnetic calorimeters”. This project 17FUN02 MetroMMC has received funding from the EMPIR programme co-financed by the Participating States and from the European Union’s Horizon 2020 research and innovation programme. Part of the calculations are performed on the JUSTUS 2 cluster supported by the state of Baden-Württemberg through bwHPC and the German Research Foundation (DFG) through grant no INST 40/575-1 FUGG (JUSTUS 2 cluster).

\newpage

\appendix
\section*{Appendix}

\section{Multi-reference full relativistic Dirac calculations of the atomic eigenstates} \label{sec:SolvingHAtom}

The atomic ground-state is calculated using a configuration interaction scheme including all 26 Fe electrons. As a one particle basis we use the Density Functional Theory Kohn-Sham orbitals, as obtained within the program FPLO \cite{koepernik1999full,opahle1999full}. The many electron basis is given by Slater determinants obtained from different possible distributions of the 26 electrons in the one particle orbital basis states. The interaction between the basis states is given by the Dirac Hamiltonian plus Breit interaction. We used the program package \textsc{Quanty} \cite{haverkort2014bands,haverkort2016quanty,brass2020ab} to calculate the ground-state, the spectra and the self-energy. In order to make the calculation tractable only determinants with the norm of the expansion coefficient larger than $10^{-6}$ are kept for the representation of the ground-state \cite{Haverkort2012}. 

The spectra are calculated using a Lanczos algorithm that generates a Krylov basis \cite{Haverkort2012}. At each step of the Lanczos algorithm the state is written as a sum over determinants. Similar to the ground-state calculations we remove from all states the determinants whose norm of the expansion coefficients is below a given threshold  of  $10^{-6}$. This assures a tractable basis size for all steps of the calculation.

\section{Multipole expansion of the light-matter interaction} \label{sec:MultipoleExpansionLightMatter}

In this section we derive an expression for the relativistic light-matter interaction in terms of magnetic and electric multipole transitions. We recall that the relativistic light-matter interaction Hamiltonian in the Coulomb gauge, using natural units with $c=\hbar=1$ is given by \cite{Johnson:51375}
\begin{eqnarray}
	\mathbf{H}_{e\gamma} = e \int \psi^{\dagger}(\mathbf{r})\boldsymbol{\alpha} \cdot \mathbf{A}(\mathbf{r}) \psi(\mathbf{r})  \mathrm{d}^3r \label{eq:RelLightMatterGeneralForm}
\end{eqnarray}
with four-component Dirac-spinor fields $\psi^{\dagger}(\mathbf{r})$ and $\psi(\mathbf{r})$ and with the matrix form of $\boldsymbol{\alpha}$,
\begin{eqnarray}
	\boldsymbol{\alpha}=\left(
	\begin{array}{cc}
		0 & \boldsymbol{\sigma} \\
		\boldsymbol{\sigma} & 0
	\end{array}\right)
\end{eqnarray}
where $\boldsymbol{\sigma}$ denote the Pauli-matrices. $\mathbf{A}(\mathbf{r})$ describes the vector potential
\begin{eqnarray}
	\mathbf{A}(\mathbf{r}) = \sum_{\mathbf{k}_{\gamma}} \sum^2_{\mu=1} \frac{1}{\sqrt{2 \omega_{\gamma}}} \big(\boldsymbol{\epsilon}_{\mu} a^{\dagger}_{\mathbf{k}_{\gamma},\mu} e^{i\mathbf{k}_{\gamma}\cdot\mathbf{r}} + \boldsymbol{\epsilon}^{\ast}_{\mu} a_{\mathbf{k}_{\gamma},\mu} e^{-i\mathbf{k}_{\gamma}\cdot\mathbf{r}}\big) \label{eq:VectorPotentialA}
\end{eqnarray}
where $\boldsymbol{\epsilon}_{\mu}$ denotes the polarization vector and $a^{\dagger}_{\mathbf{k}_{\gamma},\mu}$ and $a_{\mathbf{k}_{\gamma},\mu}$ are the creation and annihilation operators of a photon in mode $\lbrace\mathbf{k}_{\gamma}, \mu \rbrace$, respectively. Note the quantization volume is set to 1. The kinetic energy of the photons is given by
\begin{eqnarray}
	\mathbf{H}_{\gamma} = \sum_{\mathbf{k}_{\gamma}} \sum^2_{\mu=1} \omega_{\gamma} a^{\dagger}_{\mathbf{k}_{\gamma},\mu} a_{\mathbf{k}_{\gamma},\mu} \label{eq:KineticEnergyPhoton}.
\end{eqnarray}
with $\omega_{\gamma} = \vert \mathbf{k}_{\gamma} \vert = k_{\gamma}$. Inserting the vector potential (\ref{eq:VectorPotentialA}) into (\ref{eq:RelLightMatterGeneralForm}), the light-matter interaction Hamiltonian becomes
\begin{eqnarray}
	\mathbf{H}_{e \gamma} =  & e\sum_{\mathbf{k}_{\gamma}} \sum^2_{\mu=1} \frac{1}{\sqrt{2 \omega_{\gamma}}} \nonumber \\ 
	&\times \int \psi^{\dagger}(\mathbf{r}) \boldsymbol{\alpha} \cdot \left(\boldsymbol{\epsilon}_{\mu} a^{\dagger}_{\mathbf{k}_{\gamma},\mu} e^{i\mathbf{k}_{\gamma}\cdot\mathbf{r}} + \boldsymbol{\epsilon}^{\ast}_{\mu} a_{\mathbf{k}_{\gamma},\mu} e^{-i\mathbf{k}_{\gamma}\cdot\mathbf{r}} \right) \psi(\mathbf{r}) \mathrm{d}^3r \label{eq:LightMatterHamiltonianGeneralForm}. 
\end{eqnarray}
In order to simplify notation we introduce the transition operator 
\begin{eqnarray}
	\mathbf{T}_{e\gamma} =  \boldsymbol{\epsilon}_{\mu} \cdot \int \psi^{\dagger}(\mathbf{r})  \boldsymbol{\alpha} e^{i\mathbf{k}_{\gamma}\cdot\mathbf{r}} \psi(\mathbf{r}) \mathrm{d}^3r  \label{eq:PhotonTransitionOperator}
\end{eqnarray}
such that $\mathbf{H}_{e \gamma}$ takes the compact form
\begin{eqnarray}
	\mathbf{H}_{e \gamma} = e \sum_{\mathbf{k}_{\gamma}} \sum^2_{\mu = 1} \frac{1}{\sqrt{2 \omega_{\gamma}}} \left(a^{\dagger}_{\mathbf{k}_{\gamma},\mu}  \mathbf{T}_{e\gamma} + a_{\mathbf{k}_{\gamma},\mu} \mathbf{T}_{e\gamma}^{\dagger} \right) \label{eq:LightMatterHamiltonianWithT}.
\end{eqnarray}
Note that only a single photon is involved in the light-matter interaction, so $\mathbf{H}_{e \gamma}$ can induce transitions from an initial state with $n$ to a final state with $n\pm1$ photons by absorption or emission of a photon. 

In the next step, we decompose $\mathbf{T}_{e\gamma}$ into multipoles. Expanding $\boldsymbol{\epsilon}_{\mu} \cdot e^{i\mathbf{k}_{\gamma}\cdot\mathbf{r}}$ in terms of multipole potentials we obtain \cite{Johnson:51375}
\begin{eqnarray}
	\boldsymbol{\epsilon}_{\mu} \cdot e^{i\mathbf{k}_{\gamma}\cdot\mathbf{r}}  = 4 \pi \sum_{JM\lambda} i^{J-\lambda} \left(\boldsymbol{\epsilon}_{\mu} \cdot \mathbf{Y}^{(\lambda) \ast}_{JM}(\hat{k}_{\gamma})\right) \mathbf{a}^{(\lambda)}_{JM}(\mathbf{r})\label{eq:VectorPotentialSimplified}
\end{eqnarray}
with $\lambda=0,1$ and $M=-J,\dots,J$. As $\boldsymbol{\epsilon}_{\mu} \cdot e^{i\mathbf{k}_{\gamma}\cdot\mathbf{r}}$ transforms as a vector under rotations, the multipole potentials $\mathbf{a}^{(\lambda)}_{JM}(\mathbf{r})$ can be expressed as linear combinations of vector spherical harmonics $\mathbf{Y}^{(\lambda)}_{JM}(\hat{r})$:
\begin{eqnarray}
	\mathbf{a}^{(0)}_{JM}(\mathbf{r}) &= j_J(k_{\gamma}r) \mathbf{Y}^{(0)}_{JM}(\hat{r}) \nonumber\\
	\mathbf{a}^{(1)}_{JM}(\mathbf{r}) &= \left(j^{\prime}_J(k_{\gamma}r) + \frac{j_J(k_{\gamma}r)}{k_{\gamma}r}\right) \mathbf{Y}^{(1)}_{JM}(\hat{r}) \nonumber \\
	& \ \ \ \ + \sqrt{J(J+1)} \frac{j_J(k_{\gamma}r)}{k_{\gamma}r}\mathbf{Y}^{(-1)}_{JM}(\hat{r})
\end{eqnarray}
Different conventions for the vector spherical harmonics are in use. The vector spherical harmonics $\mathbf{Y}^{(\lambda)}_{JM}(\hat{r})$ noted with brackets around $\lambda$ are transverse for $\lambda=0,1$ and longitudinal for $\lambda=-1$. As such these conventions of the vector spherical harmonics are particularly suited for the expansion of the electromagnetic field, which is transverse. This comes at the cost that the transverse and longitudinal vector spherical harmonics are no longer eigenfunctions of the operator $\mathbf{L}^2$. We can define the vector spherical harmonics $\mathbf{Y}^{L}_{JM}(\hat{r})$ as eigenfunctions of the operator $\mathbf{L}^2$. The transverse and longitudinal vector spherical harmonics $\mathbf{Y}^{(\lambda)}_{JM}(\hat{r})$ are given as linear combinations of $\mathbf{Y}^{L}_{JM}(\hat{r})$, \cite{D.A.Varshalovich1988}
\begin{eqnarray}
\mathbf{Y}^{(1)}_{JM}(\hat{r}) &= \sqrt{\frac{J+1}{2J+1}} \mathbf{Y}^{J-1}_{JM}(\hat{r}) + \sqrt{\frac{J}{2J+1}} \mathbf{Y}^{J+1}_{JM}(\hat{r}) \nonumber \\
\mathbf{Y}^{(0)}_{JM}(\hat{r}) &=\mathbf{Y}^{J}_{JM}(\hat{r}) \nonumber \\
\mathbf{Y}^{(-1)}_{JM}(\hat{r}) &=\sqrt{\frac{J}{2J+1}} \mathbf{Y}^{J-1}_{JM}(\hat{r}) - \sqrt{\frac{J+1}{2J+1}} \mathbf{Y}^{J+1}_{JM}(\hat{r}) 
\end{eqnarray}
where $\mathbf{Y}^{L}_{JM}(\hat{r})$ is defined by \cite{D.A.Varshalovich1988}
\begin{eqnarray}
	\mathbf{Y}^{L}_{JM}(\hat{r}) = \sum^{L}_{m=-L} \sum^1_{q=-1} C^{JM}_{Lm1q} Y_{Lm} \mathbf{e}_q.
\end{eqnarray}
Here, $C^{JM}_{Lm1q}$ are the Clebsch-Gordan coefficients, $Y_{Lm}$ is the ordinary spherical harmonic and $\mathbf{e}_q$ the spherical basis vectors
\begin{eqnarray}
	\mathbf{e}_{+1} = -\frac{1}{\sqrt{2}} \left(\mathbf{e}_x + \mathrm{i} \mathbf{e}_y\right), \ \ \mathbf{e}_{0} = \mathbf{e}_z, \ \ \mathbf{e}_{-1} = \frac{1}{\sqrt{2}} \left(\mathbf{e}_x - \mathrm{i} \mathbf{e}_y\right).
\end{eqnarray}
In the Coulomb gauge, plane waves are transverse. Therefore, their polarization vector $\boldsymbol{\epsilon}_{\mu}$ is perpendicular to the direction of propagation, $\boldsymbol{\epsilon}_{\mu} \cdot \mathbf{k}_{\gamma} = 0$. Since the vector spherical harmonic $\mathbf{Y}^{(-1)}_{JM}(\hat{k}_{\gamma})$ is parallel to $\mathbf{k}_{\gamma}$, we find $\boldsymbol{\epsilon}_{\mu} \cdot \mathbf{Y}^{(-1)}_{JM}(\hat{k}_{\gamma}) = 0$. Thus, only multipoles with $\lambda=0,1$ contribute to the above expansion. The parts with $\lambda=0$ are referred to as \textit{magnetic} multipoles, while those with $\lambda=1$ represent the \textit{electric} multipoles \cite{Johnson:51375}. Inserting (\ref{eq:VectorPotentialSimplified}) into (\ref{eq:PhotonTransitionOperator}), the transition operator decomposed into the individual multipole moments takes the form
\begin{eqnarray}
	\mathbf{T}_{e\gamma} &= 4 \pi \sum_{JM\lambda} \mathrm{i}^{J-\lambda}  \left(\boldsymbol{\epsilon}_{\mu} \cdot \mathbf{Y}^{(\lambda) \ast}_{JM}(\hat{k}_{\gamma})\right) \nonumber \\
	 &\times \int \psi^{\dagger}(\mathbf{r})  \boldsymbol{\alpha} \cdot \mathbf{a}^{(\lambda)}_{JM}(\mathbf{r}) \psi(\mathbf{r}) \mathrm{d}^3r . \label{eq:PhotonTransitionOperatorMultipole}
\end{eqnarray}
To obtain an expression for $\mathbf{T}_{e\gamma}$ that can be used in \textsc{Quanty} \cite{haverkort2016quanty}, we express the four component Dirac-spinor field in terms of a sum of creation $e^\dagger_{\tau}$ and annihilation $e_{\tau}$ operators weighted by the single-particle wave functions $\phi^{\dagger}_{\tau}(\mathbf{r})$ and $\phi_{\tau}(\mathbf{r})$, respectively,
\begin{eqnarray}
	\label{eq:FieldOperator}
	\psi(\mathbf{r})=\sum_{\tau}\phi_\tau(\mathbf{r})e_\tau\hspace{1cm}	\psi^\dagger(\mathbf{r})=\sum_{\tau}\phi_\tau^\dagger(\mathbf{r})e^\dagger_\tau
\end{eqnarray}
with $\tau = \{n,\kappa,m\}$. Here, $n$ denotes the principal quantum number, $\kappa = (-1)^{(l+j+1/2)} \left(j + 1/2\right)$ with angular ($l$) and total angular ($j$) momentum, and $m$ the projection of $j$ on the $z$-axis. The single-particle wave functions are given by the four-component spinors
\begin{eqnarray}
	\label{eq:OrbitalWFs}
	\phi_\tau(\mathbf{r})=\left(\begin{array}{c}
		g_{n \kappa}(r)\Omega_{\kappa m}(\theta,\phi) \\
		\mathrm{i} f_{n \kappa}(r)\Omega_{\bar{\kappa} m}(\theta, \phi) \\
	\end{array}\right).
\end{eqnarray}
with $\bar{\kappa} := -\kappa$. The radial wavefunctions $G_{\tau}(r)$ and $F_{\tau}(r)$ are defined by $G_{\tau}(r) := r g_{n \kappa}(r)$ and $F_{\tau}(r) := r f_{n \kappa}(r)$. The radial functions $G_{n \kappa}(r)$ and $F_{n \kappa}(r)$ are represented on a numerical mesh and obtained from Density Functional Theory with the use of the program FPLO \cite{koepernik1999full,opahle1999full}. Upper and lower part of $\phi_\tau(\mathbf{r})$ can be written as radial part times a spherical spinor $\Omega_{\kappa m}(\theta, \phi)$. $g_{n \kappa}(r)$ is referred to as the large and $f_{n \kappa}(r)$ as the small part of the radial wave function. Inserting (\ref{eq:FieldOperator}) into (\ref{eq:PhotonTransitionOperatorMultipole}) and using (\ref{eq:OrbitalWFs}) as the single-particle wave functions, the second quantized form of the transition operator decomposed into a sum over \textit{magnetic} ($\lambda=0$) and \textit{electric} ($\lambda=1$) multipoles has the form
\begin{eqnarray}
	\mathbf{T}_{e\gamma} = \sum^1_{\lambda=0} \sum_{\tau_a\tau_b} t^{(\lambda)}_{\tau_b \tau_a}(\mathbf{k}_{\gamma}, \mu) e^\dagger_{\tau_b} e_{\tau_a} \label{eq:LightMatter2ndQuantization}
\end{eqnarray}
with matrix elements
\begin{eqnarray}
	t^{(0)}_{\tau_b \tau_a}(\mathbf{k}_{\gamma}, \mu) &= 4\pi \sum_{JM} \mathrm{i}^J \left(\boldsymbol{\epsilon}_{\mu} \cdot \mathbf{Y}^{(0) \ast}_{JM}(\hat{k}_{\gamma})\right) \int \phi^{\dagger}_{\tau_b}(\mathbf{r}) \boldsymbol{\alpha} \cdot \mathbf{a}^{(0)}_{JM}\phi_{\tau_a}(\mathbf{r}) \mathrm{d}^3r  \nonumber \\
	&= 4\pi \sum_{JM} \mathrm{i}^{J+1} \left(\boldsymbol{\epsilon}_{\mu} \cdot \mathbf{Y}^{(0) \ast}_{JM}(\hat{k}_{\gamma})\right) \nonumber \\ & \times \int \bigg( G_{\tau_b}(r) j_J(k_{\gamma}r) F_{\tau_a}(r) \langle \Omega_{\kappa_b m_b} \vert  \boldsymbol{\sigma} \cdot\mathbf{Y}^{(0)}_{JM} \vert \Omega_{\bar{\kappa}_a m_a} \rangle \nonumber \\
	& \quad\quad  - F_{\tau_b}(r) j_J(k_{\gamma}r)G_{\tau_a}(r) \langle \Omega_{\bar{\kappa}_b m_b} \vert  \boldsymbol{\sigma} \cdot \mathbf{Y}^{(0)}_{JM} \vert \Omega_{\kappa_a m_a} \rangle \bigg) \mathrm{d}r
\end{eqnarray}
and
\begin{eqnarray}
	t^{(1)}_{\tau_b \tau_a}(\mathbf{k}_{\gamma}, \mu) &= 4\pi \sum_{JM} \mathrm{i}^{J-1} \left(\boldsymbol{\epsilon}_{\mu} \cdot \mathbf{Y}^{(1) \ast}_{JM}(\hat{k}_{\gamma})\right) \int \phi^{\dagger}_{\tau_b}(\mathbf{r}) \boldsymbol{\alpha} \cdot \mathbf{a}^{(1)}_{JM}\phi_{\tau_a}(\mathbf{r}) \mathrm{d}^3r \nonumber \\
	&= 4\pi \sum_{JM} \mathrm{i}^{J}\left(\boldsymbol{\epsilon}_{\mu} \cdot \mathbf{Y}^{(1) \ast}_{JM}(\hat{k}_{\gamma})\right) \int \Bigg[ \left(j^{\prime}_J(k_{\gamma}r) + \frac{j_J(k_{\gamma}r)}{k_{\gamma}r}\right)   \nonumber \\ 
 & \quad \quad \times  \bigg( G_{\tau_b}(r) F_{\tau_a}(r) \langle \Omega_{\kappa_b m_b} \vert  \boldsymbol{\sigma} \cdot\mathbf{Y}^{(1)}_{JM} \vert \Omega_{\bar{\kappa}_a m_a} \rangle \nonumber \\ 
 & \quad \quad \quad\quad - F_{\tau_b}(r) G_{\tau_a}(r) \langle  \Omega_{\bar{\kappa}_b m_b} \vert  \boldsymbol{\sigma} \cdot\mathbf{Y}^{(1)}_{JM} \vert  \Omega_{\kappa_a m_a} \rangle \bigg) \nonumber \\
 & \quad + \frac{j_J(k_{\gamma}r)}{k_{\gamma}r}    \sqrt{J(J+1)}  \nonumber \\ 
 & \quad \quad \times \bigg(G_{\tau_b}(r) F_{\tau_a}(r)  \langle \Omega_{\kappa_b m_b} \vert  \boldsymbol{\sigma} \cdot \mathbf{Y}^{(-1)}_{JM} \vert \Omega_{\bar{\kappa}_a m_a} \rangle \nonumber \\
 &\quad \quad \quad\quad - F_{\tau_b}(r) G_{\tau_a}(r) \langle \Omega_{\bar{\kappa}_b m_b} \vert  \boldsymbol{\sigma} \cdot\ \mathbf{Y}^{(-1)}_{JM} \vert \Omega_{\kappa_a m_a} \rangle \bigg) \Bigg] \mathrm{d}r. 
\end{eqnarray}
 Both multipole operators involve matrix elements of products of the vector of Pauli matrices $\boldsymbol{\sigma}$ and vector spherical harmonics. These can be rewritten in terms of matrix elements of ordinary spherical harmonics by means of the following relations \cite{Johnson:51375}:
\begin{eqnarray}
	\langle \Omega_{\kappa_b m_b} \vert  \boldsymbol{\sigma} \cdot \mathbf{Y}^{(-1)}_{JM} \vert \Omega_{\kappa_a m_a} \rangle = -\langle  \Omega_{\bar{\kappa}_b m_b} \vert Y_{JM} \vert  \Omega_{\kappa_a m_a} \rangle ,\nonumber \\
	\langle \Omega_{\kappa_b m_b} \vert  \boldsymbol{\sigma} \cdot \mathbf{Y}^{(0)}_{JM} \vert \Omega_{\kappa_a m_a} \rangle = \frac{(\kappa_a-\kappa_b)\langle  \Omega_{\kappa_b m_b} \vert Y_{JM} \vert  \Omega_{\kappa_a m_a} \rangle}{\sqrt{J(J+1)}}, \nonumber \\
	\langle \Omega_{\kappa_b m_b} \vert  \boldsymbol{\sigma} \cdot\mathbf{Y}^{(1)}_{JM} \vert \Omega_{\kappa_a m_a} \rangle = \frac{(\kappa_a+\kappa_b) \langle \Omega_{\bar{\kappa}_b m_b} \vert Y_{JM} \vert  \Omega_{\kappa_a m_a} \rangle}{\sqrt{J(J+1)}}.
\end{eqnarray} 
Thus, the matrix element for \textit{magnetic} multipoles simplifies to
\begin{eqnarray}
	t^{(0)}_{\tau_b \tau_a} &= 4\pi \sum_{JM} \mathrm{i}^{J+1} \left(\boldsymbol{\epsilon}_{\mu} \cdot \mathbf{Y}^{(0) \ast}_{JM}(\hat{k}_{\gamma})\right) \langle \Omega_{\kappa_b m_b} \vert  Y_{JM} \vert \Omega_{\bar{\kappa}_a m_a} \rangle \nonumber \\ 
	& \ \ \ \times  \frac{-(\kappa_a+\kappa_b)}{\sqrt{J(J+1)}} R^{(0)}_{\tau_b \tau_a}(\omega_{\gamma})  \nonumber \\
	&\equiv 4\pi \sum_{JM} \mathrm{i}^{J+1} \left(\boldsymbol{\epsilon}_{\mu} \cdot \mathbf{Y}^{(0) \ast}_{JM}(\hat{k}_{\gamma})\right) \left[T^{(0)}_{JM} (\omega_{\gamma}) \right]_{\tau_b \tau_a} \label{eq:FullMMatrixElementRelativistic}
\end{eqnarray}
where the radial integral is defined by $R^{(0)}_{\tau_b \tau_a}(\omega_{\gamma}) = \int  j_J(k_{\gamma}r) \left(G_{\tau_b}(r) F_{\tau_a}(r) + F_{\tau_b}(r) G_{\tau_a}(r) \right) \mathrm{d}r$. For the \textit{electric} multipoles, we obtain matrix elements of the form
\begin{eqnarray}
	&t^{(1)}_{\tau_b \tau_a} =  4\pi \sum_{JM} \mathrm{i}^{J} \left(\boldsymbol{\epsilon}_{\mu} \cdot \mathbf{Y}^{(1) \ast}_{JM}(\hat{k}_{\gamma})\right) \langle  \Omega_{\kappa_b m_b} \vert Y_{JM} \vert  \Omega_{\kappa_a m_a} \rangle \nonumber \\
	&\ \ \times \bigg[ \frac{\kappa_b-\kappa_a}{\sqrt{J(J+1)}} R^{(1,1)}_{\tau_b \tau_a}(\omega_{\gamma}) + \sqrt{J(J+1)} R^{(1,2)}_{\tau_b \tau_a}(\omega_{\gamma}) \bigg]  \nonumber \\
	& \ \ \ \ \ \ \equiv 4\pi \sum_{JM} \mathrm{i}^{J} \left(\boldsymbol{\epsilon}_{\mu} \cdot \mathbf{Y}^{(1) \ast}_{JM}(\hat{k}_{\gamma})\right) \left[T^{(1)}_{JM} (\omega_{\gamma}) \right]_{\tau_b \tau_a}  \label{eq:FullEMatrixElementRelativistic}
\end{eqnarray}
where we have introduced the following abbreviations for the two radial integrals:
\begin{eqnarray}
	R^{(1,\alpha)}_{\tau_b \tau_a}(\omega_{\gamma}) = \begin{cases}
		\int \left(G_{\tau_b} F_{\tau_a} + F_{\tau_b} G_{\tau_a} \right)  \left(j^{\prime}_J(k_\gamma r) + \frac{j_J(k_\gamma r )}{k_\gamma r}\right) &\mathrm{for\,\,} \alpha = 1 \\
		\int \left(F_{\tau_b} G_{\tau_a} - G_{\tau_b} F_{\tau_a} \right) + \frac{j_J(k_\gamma r )}{k_\gamma r} &\mathrm{for\,\,} \alpha = 2 
	\end{cases}
\end{eqnarray}
Note that the matrix elements $\left[T^{(0)}_{JM} (\omega_{\gamma}) \right]_{\tau_b \tau_a}$ and $\left[T^{(1)}_{JM} (\omega_{\gamma}) \right]_{\tau_b \tau_a}$ include only the parts that are independent of the photon's polarization and direction of propagation. While the radial part have to be evaluated numerically, for the angular part there is an analytical solution. Matrix elements of the form $\langle \Omega_{\kappa_b m_b} \vert Y_{JM} \vert \Omega_{\kappa_a m_a} \rangle$ are conveniently evaluated by using Wigner-Eckart's theorem \cite{D.A.Varshalovich1988}:
\begin{eqnarray}
	\langle \Omega_{\kappa_b m_b} \vert Y_{JM} \vert \Omega_{\kappa_a m_a} \rangle = (-1)^{j_b-m_b} \left(\begin{array}{ccc}
		j_b & J & j_a \\
		-m_b &M & m_a \\
	\end{array}\right) 
\langle \Omega_{\kappa_b}|| Y_J ||\Omega_{\kappa_a} \rangle \label{eq:WignerEckartTheorem}
\end{eqnarray}
Here, $\langle \Omega_{\kappa_b}|| Y_J ||\Omega_{\kappa_a} \rangle$ denotes the reduced matrix element and the expression in brackets the Wigner 3j-symbol. The reduced matrix element vanishes if $l_b+J+l_a$ is odd and takes the value
\begin{eqnarray}
	\langle \Omega_{\kappa_b}|| Y_J ||\Omega_{\kappa_a} \rangle &= \sqrt{\frac{(2j_b+1)(2j_a+1)(2J+1)}{4 \pi}} \nonumber \\
	&  \times (-1)^{j_b+1/2} \left(\begin{array}{ccc}
		j_b & j_a & J \\
		-1/2 &1/2 & 0 \\
	\end{array}\right)
\end{eqnarray}
if $l_b+J+l_a$ is even. Thus, $\left[T^{(0)}_{JM}(\omega_{\gamma}) \right]_{\tau_b \tau_a}$ can be written as
\begin{eqnarray}
	\left[T^{(0)}_{JM}(\omega_{\gamma}) \right]_{\tau_b \tau_a} =& (-1)^{j_b-m_b+1} \left(\begin{array}{ccc}
		j_b & J & j_a \\
		-m_b &M & m_a \\
	\end{array}\right) \nonumber \\ 
& \times \langle \Omega_{\kappa_b}|| Y_J ||\Omega_{\kappa_a} \rangle \frac{\kappa_a+\kappa_b}{\sqrt{J(J+1)}} R^{(0)}_{\tau_b \tau_a}(\omega_{\gamma}) \label{eq:MMatrixElementRelativistic}
	\end{eqnarray}
if $l_b+J+l_a$ is odd, whereas $\left[T^{(1)}_{JM}(\omega_{\gamma}) \right]_{\tau_b \tau_a}$ becomes 
\begin{eqnarray}
	\left[T^{(1)}_{JM}(\omega_{\gamma}) \right]_{\tau_b \tau_a} & =  (-1)^{2j_b-m_b+1/2} \left(\begin{array}{ccc}
		j_b & J & j_a \\			-m_b &M & m_a \\
	\end{array}\right) \nonumber \\
& \times \left(\begin{array}{ccc}
		j_b & j_a & J \\
		-1/2 &1/2 & 0 \\
	\end{array}\right) \sqrt{\frac{(2j_b+1)(2j_a+1)(2J+1)}{4\pi}} \nonumber \\
&\times \Bigg[ \frac{\kappa_b-\kappa_a}{\sqrt{J(J+1)}}  R^{(1,1)}_{\tau_b \tau_a}(\omega_{\gamma}) + \sqrt{J(J+1)} R^{(1,2)}_{\tau_b \tau_a}(\omega_{\gamma}) \Bigg] \label{eq:EMatrixElementRelativistic}
\end{eqnarray}
if $l_b+J+l_a$ is even.

\section{Self-energy due to fluorescence decay} \label{sec:FluorescenceSelfEnergy}

The self-energy operator (\ref{eq:SelfEnergy}) can be expressed as a matrix on a basis of bound states involving no photons $\{\ket{\psi_i}\}$.  In the following, we derive explicit expressions for the matrix elements $\Sigma_{ij}(\omega) = \braket{\psi_i}{\mathbf{\Sigma}(\omega)}{\psi_j}$. For this purpose, we employ the multipole expansion of the light-matter interaction derived in \ref{sec:MultipoleExpansionLightMatter}, which allows us to  study the impact of individual multipole transitions on the spectral line-shape. As we have discussed in the main text, fluorescence decay can lead to the production of several photons. The dominant contribution, however, originates from a decay involving one additional photon for which the self-energy operator can be written as
\begin{eqnarray}
	\label{eq:SelfEnergy1Photon}
	\mathbf{\Sigma}(\omega) = \mathbf{H}^{(01)}_{e\gamma} \frac{1}{\omega - \mathbf{H}^{(11)} + \mathrm{i}0^+} \mathbf{H}^{(10)}_{e\gamma}
\end{eqnarray}
where $\mathbf{H}^{(11)} = \mathbf{H}^{(00)} + \mathbf{H}_\gamma$. Substituting (\ref{eq:LightMatterHamiltonianWithT}) for $\mathbf{H}^{(10)}_{e\gamma}$ where $\mathbf{H}^{(01)^\dagger}_{e\gamma} = \mathbf{H}^{(10)}_{e\gamma}$, the matrix elements of the self-energy operator for single-photon decays take the form
\begin{eqnarray}
	\Sigma_{i j}(\omega) = & \frac{e^2}{2}  \sum^2_{\mu= 1} \sum_{\mathbf{k}_{\gamma}} \frac{1}{\omega_{\gamma}} \nonumber \\
	 & \times  \big \langle \psi_i \big \vert \mathbf{T}_{e \gamma}^{\dagger} a_{\mathbf{k}_{\gamma},\mu} \frac{1}{\omega + \mathrm{i}0^+-\mathbf{H}^{(00)}-\mathbf{H}_{\gamma}} a^{\dagger}_{\mathbf{k}_{\gamma},\mu} \mathbf{T}_{e \gamma} \vert \psi_{j} \big \rangle.
\end{eqnarray}
The sum over all photon momenta $\mathbf{k}_{\gamma}$ can be performed by taking the continuum limit as
\begin{eqnarray}
	\sum_{\mathbf{k}_{\gamma}} \rightarrow  \frac{1}{(2\pi)^3}\int \mathrm{d}^3k_{\gamma} = \frac{1}{(2\pi)^3} \int \omega^2_{\gamma} \mathrm{d}\omega_{\gamma} \mathrm{d} \Omega_{\gamma} \label{eq:SumToIntegralPhoton}.
\end{eqnarray}
With the use of spherical coordinates the fluorescence self-energy becomes
\begin{eqnarray}
	\Sigma_{i j}(\omega) = & \frac{e^2}{2(2\pi)^3}  \sum^2_{\mu = 1} \int \omega_{\gamma} \mathrm{d}\omega_{\gamma} \mathrm{d}\Omega_{\gamma} \nonumber \\
	&\times \big \langle \psi_i \big \vert \mathbf{T}_{e \gamma}^{\dagger} a_{\mathbf{k}_{\gamma},\mu} \frac{1}{\omega + \mathrm{i}0^+-\mathbf{H}^{(00)}-\mathbf{H}_{\gamma}} a^{\dagger}_{\mathbf{k}_{\gamma},\mu} \mathbf{T}_{e \gamma} \vert \psi_{j} \big \rangle .
\end{eqnarray}
Inserting (\ref{eq:LightMatter2ndQuantization}) for the transition operator $\mathbf{T}_{e\gamma}$ with matrix elements (\ref{eq:FullMMatrixElementRelativistic}) and (\ref{eq:FullEMatrixElementRelativistic}) and further using that $\mathbf{H}_{\gamma} a^{\dagger}_{\mathbf{k}_{\gamma}, \mu} \vert \psi_i \rangle = \omega_{\gamma} a^{\dagger}_{\mathbf{k}_{\gamma}, \mu} \vert \psi_i \rangle$, we finally obtain
\begin{eqnarray}
	\Sigma_{i j}(\omega) = & 4 \alpha \sum^2_{\mu=1} \int \sum_{\tau_a\tau_b\tau^\prime_a \tau^\prime_b} \sum_{JM\lambda} \sum_{J^\prime M^\prime \lambda^\prime} \left(\mathbf{Y}^{(\lambda^{\prime})}_{J^{\prime}M^{\prime}}(\hat{k}_{\gamma}) \cdot \boldsymbol{\epsilon}^{\ast}_{\mu}\right) \left(\boldsymbol{\epsilon}_{\mu} \cdot \mathbf{Y}^{(\lambda) \ast}_{J M}(\hat{k}_{\gamma}) \right) \nonumber \\ 
 & \ \ \times    \left[T^{(\lambda^\prime)}_{J^\prime M^\prime}(\omega_{\gamma})\right]_{\tau^\prime_b \tau^\prime_a} \left[T^{(\lambda)}_{JM}(\omega_{\gamma})\right]_{\tau_b \tau_a} \nonumber \\
	&\ \ \times  \big \langle \psi_i \big \vert e^\dagger_{\tau^\prime_b} e_{\tau^\prime_a} \frac{1}{\omega + \mathrm{i}0^+-\mathbf{H}^{(00)}-\omega_{\gamma}}  e^\dagger_{\tau_b} e_{\tau_a}\big \vert \psi_{j} \big \rangle \omega_{\gamma} \mathrm{d}\omega_{\gamma} \mathrm{d}\Omega_{\gamma}
\end{eqnarray}
with fine-structure constant $\alpha$. As the vector spherical harmonics with $\lambda=0,1$ both are orthogonal to $\hat{k}_{\gamma}$, the sum over the two polarizations can be performed as \cite{Johnson:51375}
\begin{eqnarray}
	\sum^2_{\mu=1} \left(\mathbf{Y}^{(\lambda^{\prime})}_{J^{\prime}M^{\prime}}(\hat{k}_{\gamma}) \cdot \boldsymbol{\epsilon}^{\ast}_{\mu}\right) \left(\boldsymbol{\epsilon}_{\mu} \cdot \mathbf{Y}^{(\lambda) \ast}_{J^{}M^{}}(\hat{k}_{\gamma}) \right) = \left(\mathbf{Y}^{(\lambda^{\prime})}_{J^{\prime}M^{\prime}}(\hat{k}_{\gamma}) \cdot\mathbf{Y}^{(\lambda) \ast}_{J^{}M^{}}(\hat{k}_{\gamma})\right) \label{eq:IdentityVecSpher-Times-PolVec}
\end{eqnarray}
and  the self-energy simplifies to
\begin{eqnarray}
	\Sigma_{i j}(\omega)  = & 4 \alpha \int \sum_{\tau_a\tau_b\tau^\prime_a \tau^\prime_b} \sum_{JM\lambda} \sum_{J^\prime M^\prime \lambda^\prime}  \left(\mathbf{Y}^{(\lambda^{\prime})}_{J^{\prime}M^{\prime}}(\hat{k}_{\gamma}) \cdot\mathbf{Y}^{(\lambda) \ast}_{J^{}M^{}}(\hat{k}_{\gamma})\right) \nonumber \\
 & \ \ \times \left[T^{(\lambda^\prime)}_{J^\prime M^\prime}(\omega_{\gamma})\right]_{\tau^\prime_b \tau^\prime_a}   \left[T^{(\lambda)}_{JM}(\omega_{\gamma})\right]_{\tau_b \tau_a} \nonumber \\
	& \ \  \times \big \langle \psi_i \big \vert e^\dagger_{\tau^\prime_b} e_{\tau^\prime_a}  \frac{1}{\omega + \mathrm{i}0^+-\mathbf{H}^{(00)}-\omega_{\gamma}}  e^\dagger_{\tau_b} e_{\tau_a} \big \vert \psi_{j} \big \rangle \omega_{\gamma} \mathrm{d}\omega_{\gamma} \mathrm{d}\Omega_{\gamma}.
\end{eqnarray}
To evaluate the integral over the solid angle, we exploit that the vector spherical harmonics obey the orthonormality relation \cite{D.A.Varshalovich1988}
\begin{eqnarray}
	\int \mathbf{Y}^{(\lambda^{\prime}) \ast}_{J^{\prime}M^{\prime}}(\hat{k}_{\gamma}) \cdot \mathbf{Y}^{(\lambda)}_{JM}(\hat{k}_{\gamma}) \mathrm{d}\Omega_{\gamma} = \delta_{J^{\prime}J}\delta_{M^{\prime}M}\delta_{\lambda^{\prime}\lambda} \label{eq:OrthonomalityVectorSphericalHarmonics}
\end{eqnarray}
which enables us to directly integrate over the photon's momentum directions: 
\begin{eqnarray}
	\Sigma_{i j}(\omega) = & 4 \alpha \int \sum_{\tau_a\tau_b\tau^\prime_a \tau^\prime_b} \sum_{JM\lambda}  \left[T^{(\lambda)}_{JM}(\omega_{\gamma})\right]_{\tau^\prime_b \tau^\prime_a} \left[T^{(\lambda)}_{JM}(\omega_{\gamma})\right]_{\tau_b \tau_a} \nonumber \\
	& \ \ \ \times \big \langle \psi_i \big \vert e^\dagger_{\tau^\prime_b} e_{\tau^\prime_a} \frac{1}{\omega + \mathrm{i}0^+-\mathbf{H}^{(00)}-\omega_{\gamma}}  e^\dagger_{\tau_b} e_{\tau_a} \big \vert \psi_{j} \big \rangle \omega_{\gamma} \mathrm{d}\omega_{\gamma} 
\end{eqnarray}
Note that different multipoles as well as magnetic and electric transitions do not interfere. To evaluate the integral over $\omega_{\gamma}$, we replace
\begin{eqnarray}
	\frac{1}{\omega + \mathrm{i}0^+-\mathbf{H}^{(00)}-\omega_{\gamma}} \longrightarrow \sum_{n} \frac{\vert \psi_n \rangle \langle \psi_n \vert }{\omega+\mathrm{i}0^+-E_n-\omega_{\gamma}} \nonumber
\end{eqnarray}
where $\vert \psi_n \rangle$ are eigenstates of $\mathbf{H}^{(00)}$ with eigenenergies $E_n$ measured from the $^{55}\mathrm{Mn}$ ground state. Then, we can rewrite $\frac{1}{\omega+\mathrm{i}0^+-E_n-\omega_{\gamma}}$ as 
\begin{eqnarray}
	\frac{1}{\omega+\mathrm{i}0^+-E_n-\omega_{\gamma}}= \mathcal{P} \frac{1}{\omega-E_n-\omega_{\gamma}}- i\pi\delta(\omega-E_n-\omega_{\gamma})
\end{eqnarray}
where $\mathcal{P}$ represents the Cauchy principal value. Now the self-energy is split into a real and an imaginary part, both involving an integral over $\omega_{\gamma}$. With this, the latter is given by
\begin{eqnarray}
	\mathrm{Im} \Sigma_{ij} (\omega)  = & -4 \pi \alpha \sum_{\tau_a\tau_b\tau^\prime_a \tau^\prime_b} \sum_{JM\lambda} \sum_{n}  \left[T^{(\lambda)}_{JM}(\omega-E_n)\right]_{\tau^\prime_b \tau^\prime_a} \left[T^{(\lambda)}_{JM}(\omega-E_n)\right]_{\tau_{b} \tau_{a}} \nonumber \\ & 
 \times \left(\omega-E_n \right)   \big \langle \psi_i \big \vert e^\dagger_{\tau^\prime_b} e_{\tau^\prime_a}  \big \vert \psi_n \big \rangle \big \langle \psi_n \big \vert e^\dagger_{\tau_b} e_{\tau_a} \big \vert \psi_{j} \big \rangle  \Theta(\omega-E_n) \label{eq:ImPartSelfenergy}.
\end{eqnarray} 

The real part $\mathrm{Re} \Sigma_{ij} (\omega)$, on the other hand, is related to the imaginary part $\mathrm{Im} \Sigma_{ij} (\omega)$ via the Kramers-Kronig relations:
\begin{eqnarray}
	\mathrm{Re} \Sigma_{ij} (\omega) = \frac{2}{\pi} \mathcal{P} \int_{0}^{\infty} \frac{\omega^{\prime} \mathrm{Im} \Sigma_{ij} (\omega^{\prime})}{\omega^{\prime 2} - \omega^2} \ \mathrm{d}\omega^{\prime} \label{eq:RealPartFYSEKramersKronig}
\end{eqnarray}
At first glance, it does not look like we have simplified the calculation of the real part by applying the Kramers-Kronig relations, since we have transformed one integral into another one, both of which extend to infinity. However, (\ref{eq:RealPartFYSEKramersKronig}) has the advantage that the integrand decays quickly, so in practice one can restrict the integration domain to a finite interval. The disadvantage here is that this interval exceeds the energy window on which the self-energy is needed, so it is necessary to determine $\mathrm{Im} \Sigma_{ij} (\omega)$ on this energy domain as well.	

\newpage

\section*{Refferences}

\end{document}